\documentclass[pdflatex,sn-nature]{sn-jnl}

\usepackage{graphicx}%
\usepackage{multirow}%
\usepackage{amsmath,amssymb,amsfonts}%
\usepackage{amsthm}%
\usepackage{mathrsfs}%
\usepackage[title]{appendix}%
\usepackage{xcolor}%
\usepackage{textcomp}%
\usepackage{manyfoot}%
\usepackage{booktabs}%
\usepackage{algorithm}%
\usepackage{algorithmicx}%
\usepackage{algpseudocode}%
\usepackage{listings}%
\usepackage{makecell} 

\newcounter{suppfig}

\theoremstyle{thmstyleone}%
\theoremstyle{thmstyletwo}%

\theoremstyle{thmstylethree}%

\begin{document}

\title[Article Title]{Learning to Mitigate Post-Outage Load Surges: A Data-Driven Framework for Electrifying and Decarbonizing Grids}


\author[1]{Wenlong Shi}
\author[1]{Dingwei Wang}
\author[1]{Liming Liu}
\author[1,*]{Zhaoyu Wang}
\affil[1]{Iowa State University, Department of Electrical and Computer Engineering, Ames, IA, USA}

\affil[*]{Corresponding Author: Zhaoyu Wang (wzy@iastate.edu)}


\abstract{Electrification and decarbonization are transforming power system demand and recovery dynamics, yet their implications for post-outage load surges remain poorly understood. Here we analyze a metropolitan-scale heterogeneous dataset for Indianapolis comprising 30,046 feeder-level outages between 2020 and 2024, linked to smart meters and submetering, to quantify the causal impact of electric vehicles (EVs), heat pumps (HPs) and distributed energy resources (DERs) on restoration surges. Statistical analysis and causal forest inference demonstrate that rising penetrations of all three assets significantly increase surge ratios, with effects strongly modulated by restoration timing, outage duration and weather conditions. We develop a component-aware multi-task Transformer estimator that disaggregates EV, HP and DER contributions, and apply it to project historical outages under counterfactual 2035 adoption pathways. In a policy-aligned pathway, evening restorations emerge as the binding reliability constraint, with exceedance probabilities of 0.057 when 30\% of system load is restored within the first 15 minutes. Mitigation measures, probabilistic EV restarts, short thermostat offsets and accelerated DER reconnection, reduce exceedance to 0.019 and eliminate it entirely when 20\% or less of system load is restored. These results demonstrate that transition-era surges are asset-driven and causally linked to electrification and decarbonization, but can be effectively managed through integrated operational strategies.
}




\maketitle


Electrification and decarbonization are rapidly transforming end-use demand across advanced economies. Electric vehicle (EV) registrations have grown at double-digit rates in recent years, with national roadmaps in the United States and China targeting 30–60\% market shares by 2035 \cite{IEAEV}. Policy incentives are accelerating the uptake of electric heat pumps (HPs), and annual HP sales already exceed gas boilers in France, Germany and Italy \cite{IEAHP}. Distributed energy resources (DERs) are also expanding: the European Union’s REPowerEU strategy calls for at least 700~GW of solar photovoltaics by 2030 \cite{EUDER}. While electrification supports climate goals \cite{woody2023decarbonization,rosenow2022heating} and decarbonization has been linked to grid resilience enhancement under extreme weather \cite{zhao2024impacts}, these trends introduce restoration dynamics dominated by EVs, HPs and DERs whose joint, time-coupled post-outage behavior remains poorly quantified at system scale.

Post-outage load surge has traditionally been modeled under the framework of cold load pickup (CLPU), driven by diversity loss in thermostatically controlled loads and coincident appliance restarts \cite{wang2023sequential}. However, rising levels of electrification and decarbonization are introducing entirely new surge drivers. EV charging may exceed feeder capacity when large numbers of chargers resume operation simultaneously \cite{li2024impact}. Heat pumps may exhibit elevated demand beyond their steady-state levels as they work to restore indoor temperature setpoints \cite{NRELHP}. Meanwhile, DERs, rather than providing support, can temporarily withdraw from the system due to anti-islanding protocols, which enforce delayed reconnection following voltage recovery \cite{photovoltaics2018ieee}. These dynamics collectively reshape the character of post-outage loading and challenge grid resilience.

Resilience concerns are increasingly acute under climate-amplified hazards \cite{kulkarni2024enhancing}. A key operational constraint during restoration is available headroom, as excessive surges can exhaust operating reserves, inducing instability, and compromising re-energization sequences \cite{xu2025quantifying}. During Winter Storm Uri (2021), prolonged subfreezing temperatures in Texas led to exceptional reliance on electric heating, illustrating how synchronized HP recovery and strip-heat activation can produce large restoration spikes and hinder recovery \cite{lee2022community}. Prior resilience research emphasizes risk assessment, long-term hardening and emergency operations, often using historical outage records and simulation-based models \cite{feng2025hurricane,feng2022tropical,wu2022fragmentation,ji2016large}, and promotes both infrastructure reinforcement and DER integration \cite{jain2017data,sturmer2024increasing,bennett2021extending}, alongside optimal dispatch and adaptive reconfiguration to accelerate service restoration \cite{choobdari2024robust,jacob2024real}. However, metropolitan-scale, asset-level quantification of transition-era post-outage surge mechanisms, and their causal links to EV/HP/DER penetration, remains under-presented. In headroom limited scenarios, even well-designed restoration strategies may fail unless surge behavior is explicitly estimated and managed.

Recent metering and control deployments now make such estimation feasible. Advanced metering infrastructure (AMI) records 15-minute interval demand for individual customers across most North American and European utilities \cite{AMIDOE}. Outage Management Systems (OMSs) provide event timing and affected-customer data \cite{FLISRDOE}. Managed-charging and thermostat-incentive programs add submeters for EV supply equipment and HP circuits, while Distributed Energy Resource Management Systems (DERMSs) track installed capacities and operating states. Together, these metropolitan-scale, asset-level data streams enable causal analysis and learning-based estimation of asset-specific surge contributions, extending beyond conventional CLPU.

In this study, we examine how rising electrification and decarbonization reshape post-outage load surges using a metropolitan-scale dataset for Indianapolis. The dataset comprises 30,046 feeder-level outage events recorded between 2020 and 2024, linked to 15-minute smart meter data for roughly 520,000 customers and supplemented with high-resolution EV and HP submeters. Statistical analysis and causal forest inference confirm that increasing EV, HP, and DER penetrations causally elevate surge ratios, with effects strongly modulated by restoration timing, outage duration, and ambient conditions. Building on these insights, a component-aware multi-task Transformer estimator is trained to capture asset-specific surge contributions and applied to project historical outages under counterfactual 2035 adoption trajectories. Results show that without intervention, post-outage surges exceed available headroom during evening restorations under the policy-aligned pathway, with exceedance probabilities reaching 0.057 at a restoration scale of $\alpha=0.30$. With mitigation, combining probabilistic EV restarts, brief thermostat offsets, and accelerated DER reconnection, exceedance declines to 0.019 at $\alpha=0.30$ and is eliminated entirely for $\alpha\leq0.20$, thereby expanding the range of restoration scales that can be safely accommodated within system headroom.

\begin{figure}[t]
\centering
\includegraphics[width=1.0\textwidth]{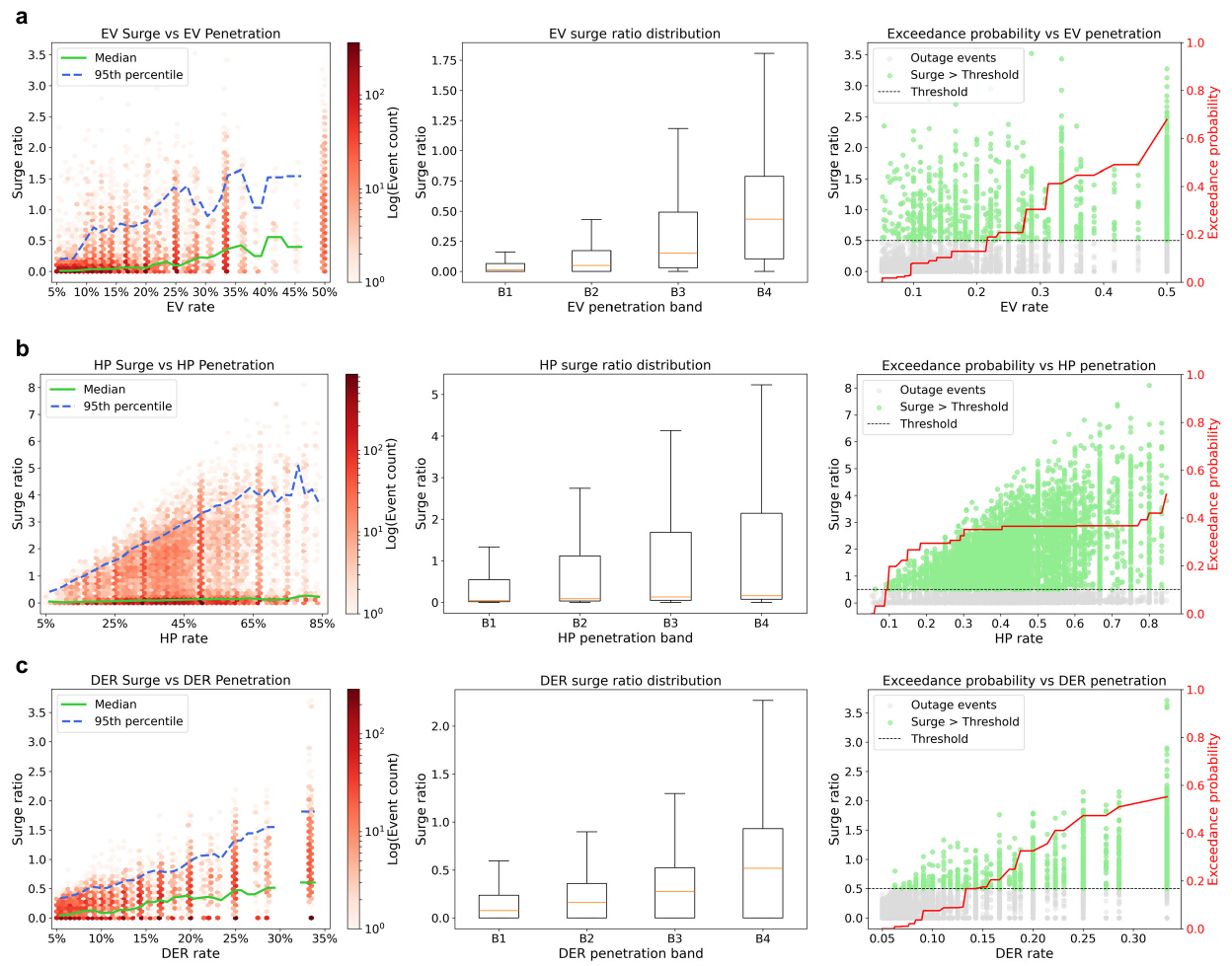}
\caption{\textbf{Statistical analysis of post-outage surge ratios under EV, HP, and DER penetrations.} (a) EV surge characterization: left, hexbin density of outage events with median (green) and 95th percentile (blue dashed) surge ratios versus EV penetration; middle, boxplots of four penetration bands (5–10\%, 10–20\%, 20–35\%, 35–50\%); right, exceedance probability curve showing the fraction of events above the threshold (black dashed) with estimated probability (red). (b) HP surge characterization: left, hexbin plots with median and 95th percentile; middle, boxplots for four bands (5–25\%, 25–45\%, 45–65\%, 65–85\%); right, exceedance probability curve. (c) DER surge characterization: left, hexbin plots with median and 95th percentile; middle, boxplots for four bands (5–10\%, 10–15\%, 15–25\%, 25–35\%); right, exceedance probability curve.}\label{fig1}
\end{figure}

\section*{Post-outage load surge assessment}

In this section, we explore how electrification and decarbonization, manifested in rising penetration of assets,  including EVs, HPs and DERs, jointly drives post-outage load surges. Our analysis draws on $30,046$ feeder-level outage events recorded between 2020 and 2025 for the Indianapolis metropolitan service area, USA. For each event, we identify the interrupted customers, extract their 15-minute interval AMI meter readings, and calculate the corresponding asset surge ratio (see Supplementary Note 1). Event-level asset penetration rates are derived from utility rebate program enrollments and the DERMS. To ensure comparability over time, all the metrics are normalized to remove differences in population and system growth. The event-level asset surge ratios and asset penetration rates are defined to perform the study (Methods).

Fig. \ref{fig1}(a) presents the asset surge ratios against penetration rates, with green curves denoting the mean and blue curves the 95th percentile. Fig. \ref{fig1}(b) summarizes the distribution of events across four penetration bands, from B1 (lowest percentage) to B4 (highest percentage), for EVs, HPs, and DERs. Across all three assets, the mean surge ratio increases along with penetration. By contrast, the 95th-percentile curves reveal a sharper escalation, highlighting the behavior of the most severe 5\% of events. For example, the 95th-percentile for HPs rises from 1.64 at 25\% penetration to 2.91 by 45\%, indicating extreme surges intensify as penetration increases. Nevertheless, since the majority of outages occur in low penetration bands, high penetration events remain relatively sparse. In other words, the probability of encountering such events is low, despite their elevated surge risk. To quantify this tail risk, Fig. \ref{fig1}(c) plots the exceedance probability curves for fixed thresholds (Supplementary Note 1). Across all three assets, exceedance probabilities increase monotonically with penetration and rise more rapidly once penetration enters the higher bands, indicating that high‐penetration outages disproportionately populate the upper tail. For example, DER  exceedance probability is around zero below 8\% penetration, increases to 0.2 near 15\%, and climbs to nearly 0.6 by 30\%. These observations confirm that, despite their lower empirical frequency, high-penetration outages are associated with a significantly increased likelihood of large post-outage surges.

\begin{figure}[h]
\centering
\includegraphics[width=0.80\textwidth]{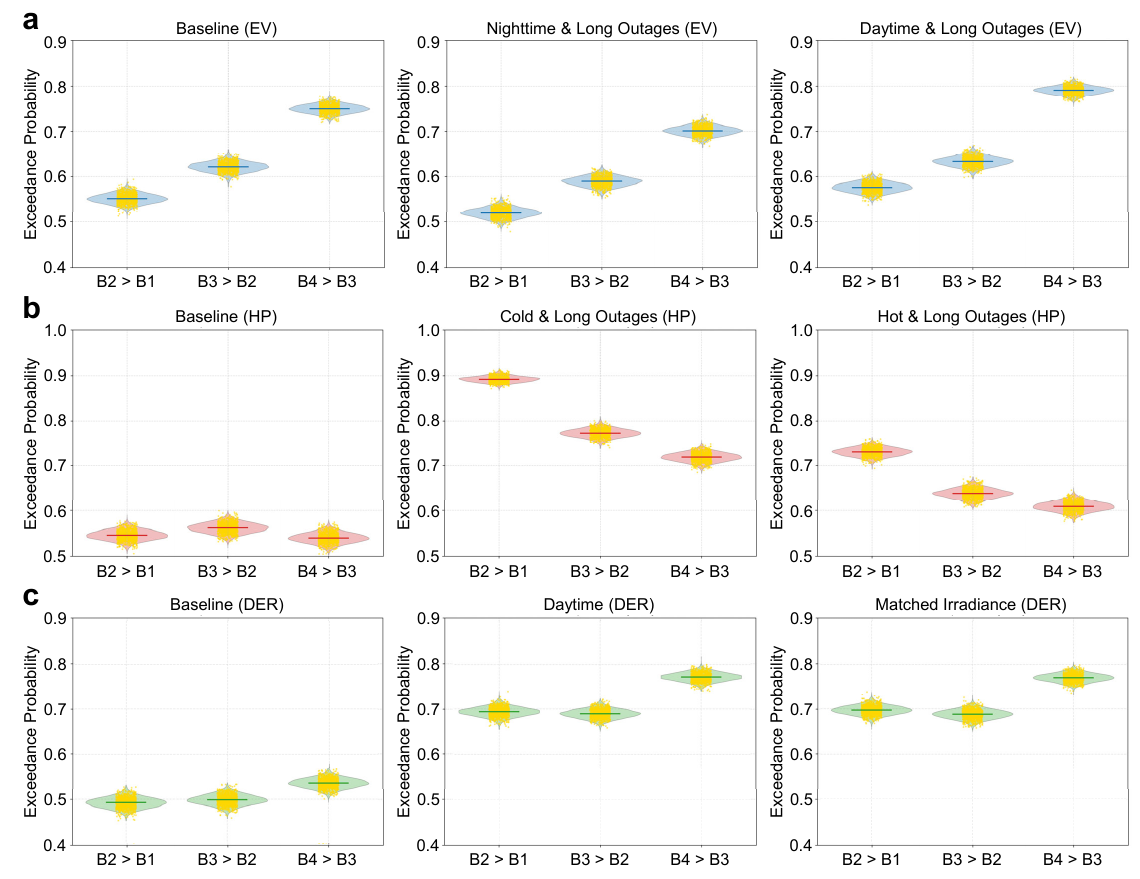}
\caption{\textbf{Exceedance probabilities of surge ratios across penetration bands.} Violin plots show the distribution of exceedance probabilities, yellow dots indicate individual bootstrap estimates, and the horizontal line marks their mean. Each row corresponds to an asset: (a) EVs, (b) HPs, and (c) DERs. Unconditional baseline panels include all events. Subset panels restrict comparisons to events with comparable contextual conditions: EVs (Nighttime \& Long Outages: the timing of restoration $\leq 6$ or $\geq 18$, duration $\geq 4$h and Daytime \& Long Outages: the timing of restoration 8–16h, duration $\geq 4$h); HP (Cold \& Long Outages: temperature $\leq$ 50°F, duration $\geq$ 4h and Hot \& Long Outages: temperature $\geq$ 80°F, duration $\geq$ 4h); DERs (Daytime: GHI $\geq$ 200 W/m² and Matched Irradiance: pairs drawn within GHI bins  [200,400], [400,700], [700,1000], and [1000,1400] W/m²).}\label{fig3}
\end{figure}

To further assess the  penetration–surge relationship, we perform a bootstrap analysis to estimate the probability that the surge ratio of an event from a higher penetration band exceeds that of an event from a lower band  (Supplementary Note 2). When pooling all events, the resulting probabilities for each asset cluster around 0.5, reflecting the confounding influence of heterogeneous conditions that diminish the observable impact of penetration (Fig. \ref{fig3}). To address this limitation, we conduct bootstrap analyses restricted to comparable subsets of events with comparable characteristics. Under these controls, the exceedance probabilities increase significantly, typically to approximate 0.7, thereby revealing a consistent positive association between penetration level and surge ratio. These results indicate that penetration is a necessary but not sufficient driver of post-outage surges: for EVs, the restoration timing and outage duration substantially affect outcomes; for HPs, ambient temperature and outage duration exert dominant influence; and for DERs, solar irradiance and time of day are critical modifiers. In addition, we conduct hypothesis testing across all three assets (Supplementary Note 2). It shows that adjacent penetration band comparisons with statistically significant differences ($p<0.01$) always show the higher penetration band exhibiting a larger surge ratio. To ensure our findings are not driven by heavy tailed surges, we also examine three thresholds corresponding to the 70th, 80th, and 90th percentiles of the surge ratio distribution (Supplementary Note 2). For each asset and at each threshold, the probability of outage events whose surge ratio falls below the threshold decreases steadily as penetration rises from B1 to B4. These downward trends indicate that the observed increases in median values
and upper-tail exceedance probabilities persist even when the analysis is focusing on
the central bulk of the distribution.

\begin{figure}[h]
	\centering
	\includegraphics[width=0.5\textwidth]{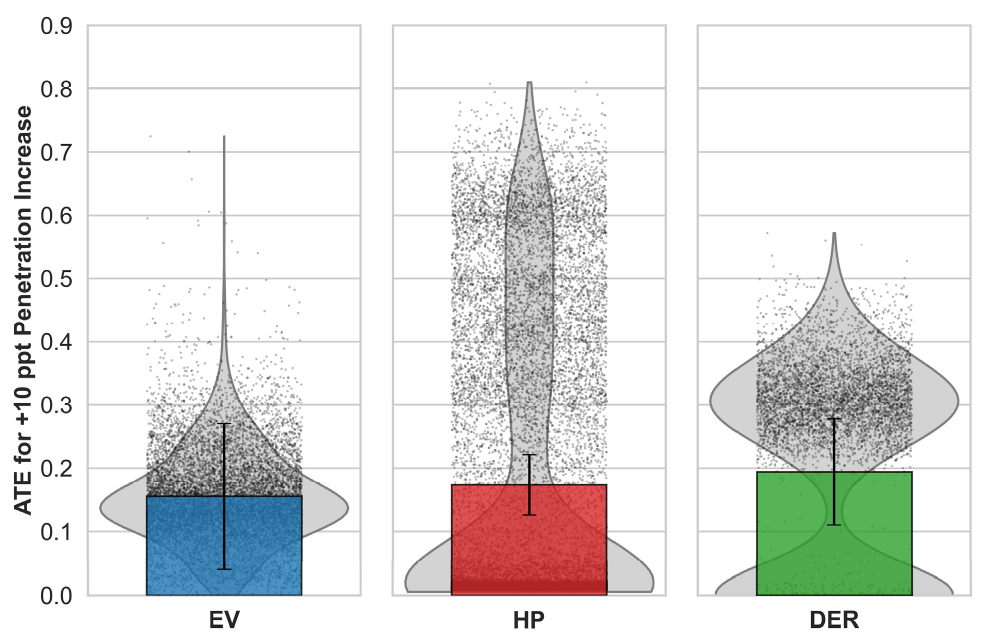}
	\caption{\textbf{Distribution of estimated ATEs for a 10 percentage-point increase in asset penetration.} Each panel shows causal forest estimates of the effect of a +10 percentage-point increase in EV, HP, or DER penetration on the surge ratio. The gray violin depicts the distribution of local ATE across outage events, and black dots indicate individual estimates. The colored bar marks the mean of these local effects, while the vertical error bar shows the empirical 95\% interval of the local effects.}\label{fig4}
\end{figure}

To evaluate whether the observed trend is confounded by exogenous factors such as extreme weather or grid evolution, we estimate the causal effect of EV, HP, and DER penetration on the surge ratio using a causal forest. We consider ambient temperature, outage duration, number of customers affected, and time-of-day, with solar irradiance additionally included for DERs as confounders (Methods). This approach isolates the marginal effect of asset penetration on post-restoration surges by learning treatment-effect heterogeneity across events while controlling for confounders. Fig. \ref{fig4} illustrates the distribution of local average treatment effects (ATEs), from which we compute the ATE, for a +10 percentage-point increase in penetration. All three assets exhibit positive and statistically significant mean ATEs: EVs +0.155 (95\% CI [0.039, 0.271]), HPs +0.173 (95\% CI [0.127, 0.219]), and DERs +0.194 (95\% CI [0.106, 0.282]), with confidence intervals excluding zero, confirming robust causal effects. The violin shapes further reveal differences in heterogeneity across assets. For EVs, the distribution is relatively compact and unimodal, indicating a consistent causal effect across most events. HPs exhibit a taller violin with a narrow central concentration and a long upper tail, reflecting greater heterogeneity. In other words, many events exhibit moderate effects, while some show much stronger surges, consistent with the sensitivity of heating and cooling demand to temperature and outage duration. DERs present a broader and more symmetric distribution after controlling for solar irradiance, suggesting moderate variability depending on restoration timing relative to solar availability. Together, these results demonstrate that higher penetration of all three assets causally increases post-restoration surges.

\section*{Component-aware load surge estimator}

A multi-task Transformer is trained end-to-end for estimating four post-outage surge components: EV charging, HP reactivation, DER delayed reconnection, and a residual category that includes CLPU (Method). The training dataset is constructed using submeter data for EV and HP loads, and DER reconnection profiles from industry-standard delay models (Supplementary Note 1). Model inputs consist of outage timestamp embeddings (hour, day-of-week, month), weather conditions (temperature, global horizontal irradiance, precipitable water), and penetration rates of EVs, HPs, and DERs. These inputs are processed through a shared Transformer encoder, followed by global average pooling, and then passed to four parallel decoder multilayer perceptrons (MLPs) heads, each corresponding to a load surge component. The model outputs four predicted incremental surge ratios, one for each component. On the 20\% testing dataset of 30,046 events, the component-aware surge estimator demonstrates strong predictive performance, achieving coefficient of determination ($R^2$) values of 0.864 for EVs, 0.917 for HPs, 0.861 for DER and 0.782 for the residual component. In absolute terms, the EV head yields a root-mean-square error (RMSE) of 0.288 and a mean absolute error (MAE) of 0.138. For HPs, the RMSE and MAE are 0.369 and 0.176, respectively; for DERs, 0.324 and 0.133; and for the residual component, 0.076 and 0.061. These results demonstrate that the model effectively learns the physical dependencies driving each surge component.

\begin{figure}[t]
	\centering
	\includegraphics[width=1\textwidth]{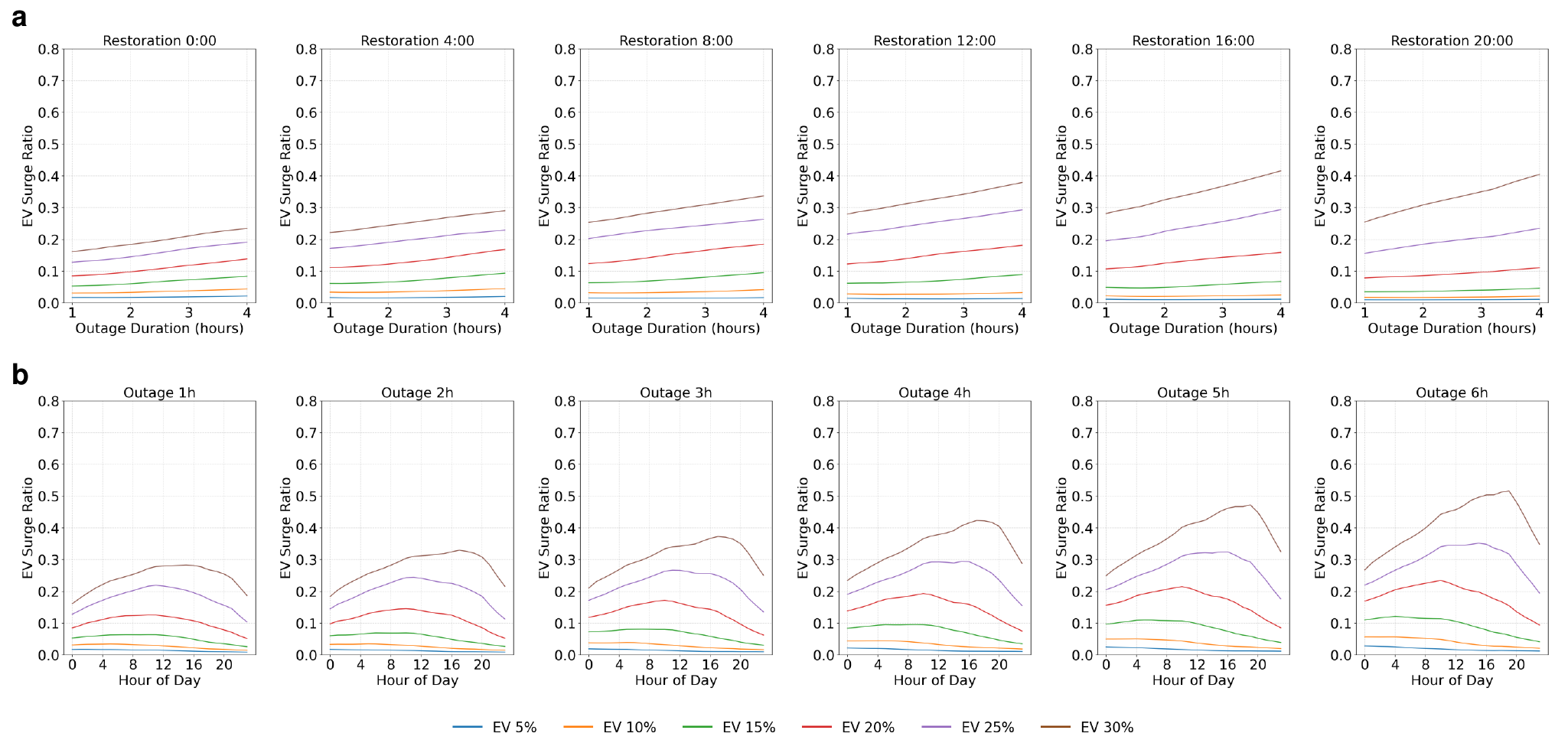}
	\caption{\textbf{Parametric analyses of EV-induced surge ratios under varying outage durations and restoration times.}  (a) EV surge ratios as functions of outage duration (1–4 h) across six representative restoration times (0:00, 4:00, 8:00, 12:00, 16:00, and 20:00). (b) EV surge ratios as functions of restoration time (0–23 h) for six representative outage durations (1–6 h).}\label{EVEST}
\end{figure}

For EV induced surges, we conduct two parametric analyses: (i) varying the restoration time across the diurnal cycle at fixed outage durations, and (ii) varying the outage duration at fixed restoration times, with all other covariates held at reference values (Fig. \ref{EVEST}). Across both analyses, the model produces a consistent diurnal structure characterized by a dominant evening maximum (18:00-21:00) and a secondary midday feature, while overnight surges remain modest. At 30\% EV penetration, the evening surge increases from approximately 0.27-0.30 for a 1 h outage to 0.48-0.52 for 5-6 h outages, and at 25\% penetration, the evening peak reaches 0.30-0.35, whereas at $\leq$10\% penetration it remains below 0.12 even for extended outages. This temporal dependence reflects the loss of charging diversity during outages and the subsequent synchronized resumption of charging upon power restoration. The outage-duration analysis further demonstrates that extended interruptions allow greater accumulation of deferred charging demand, resulting in more pronounced surges. The growth with duration is monotonic and becomes substantial once EV penetration exceeds 20–25\%. Another finding is the strong modulation of surge magnitude by restoration timing. For fixed outage duration and penetration, evening restorations consistently result in the largest surges, whereas midnight and early-morning restorations yield the smallest. This indicates that surge behavior is governed by the joint influence of penetration, duration, and timing, rather than any single factor in isolation. For instance, surge ratios $\geq$0.30 typically occur only when penetration exceeds 25\% and outages of at least 3-4 h are restored in the evening, whereas short outages or midday restorations remain comparatively limited even under high adoption scenarios. These findings suggest that evening restorations at penetrations above 25\% and durations beyond 3 h represent the most critical stress scenario for system operators.

\begin{figure}[t]
\centering
\includegraphics[width=1.0\textwidth]{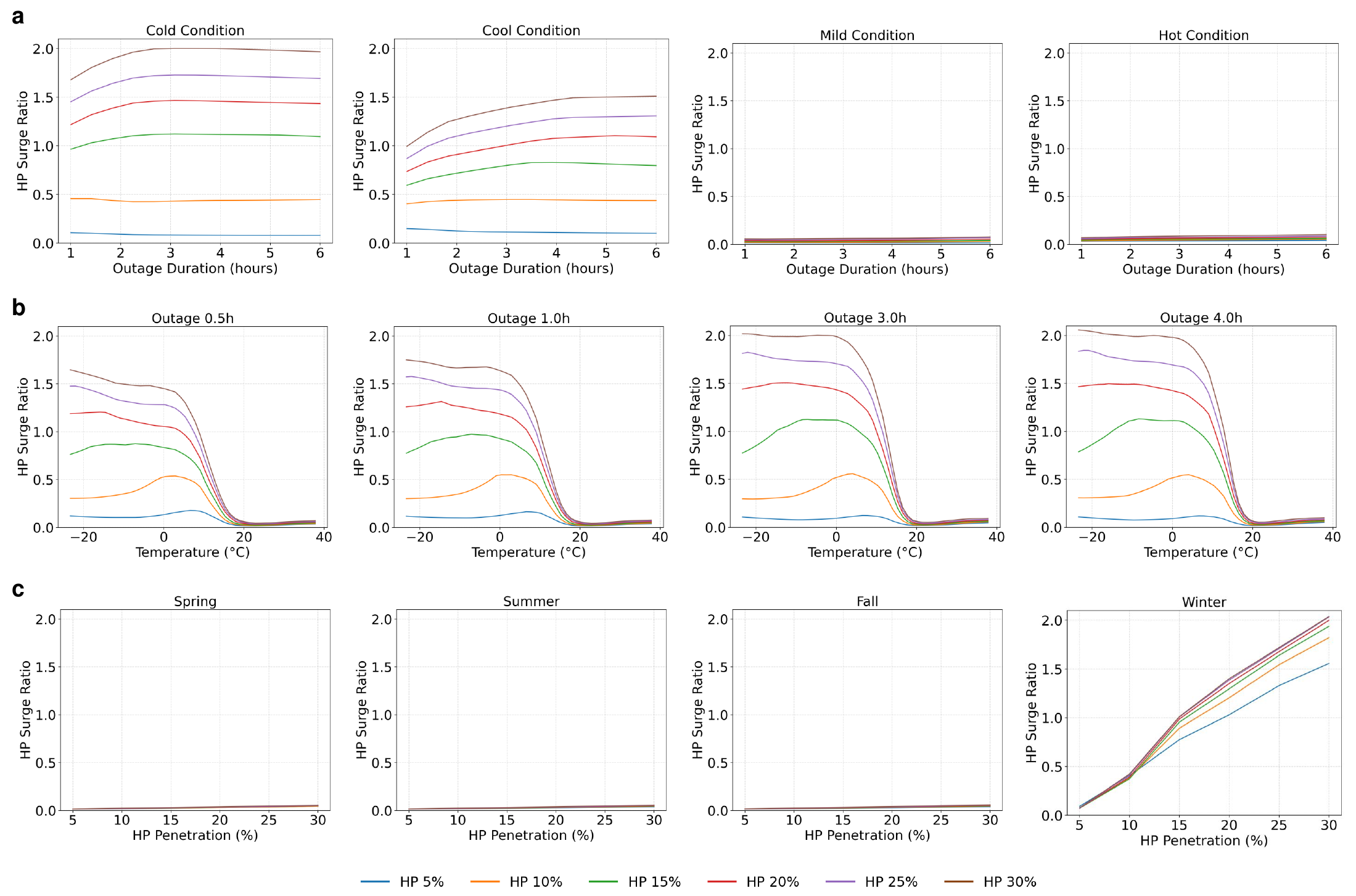}
\caption{\textbf{Parametric analyses of HP-induced surge ratios under varying outage durations, ambient temperatures, and seasonal conditions.} (a) HP surge ratios as functions of outage duration (1–6 h) across four representative temperature regimes (cold, cool, mild, hot). (b) HP surge ratios as functions of ambient temperature (–20 °C to 40 °C) for four representative outage durations (0.5–4 h). (c) HP surge ratios as functions of  penetration across four seasons. }\label{HPEST}
\end{figure}

HP induced surges exhibit different dynamics from EVs, reflecting the thermodynamic coupling of heating demand to ambient temperature (Fig. \ref{HPEST}). Across outage durations, surge ratios display a strong threshold behavior. Above 15 °C, surge ratio remain negligible (under 0.1 even at 30\% penetration), indicating that resynchronized cooling loads contribute little to restoration demand. Under cool condition (5–15 °C), surge ratio decline steeply as duty cycles diminish, but below 5 °C the response escalates rapidly, reaching 1.2–1.5 at 20–25\% penetration and exceeding 2.0 at 30\% penetration under very cold conditions. This sharp escalation reflects not only compressor synchronization but also the widespread activation of auxiliary resistance strip heating. Outage duration analyses reinforce these findings. In mild or hot conditions, surge ratios remain $<$0.1 across penetrations even for 6 h events. In cool conditions, surge ratios grow steadily with duration, exceeding 1.4 at 30\% penetration after 6 h outages, consistent with the accumulation of unmet compressor demand. Under very cold conditions, however, growth saturates quickly. Surge ratios surpass 1.8 after only 2 h interruptions and plateau thereafter, suggesting that once nearly all loads have transitioned to strip heating, additional outage duration contributes little incremental synchronization. Seasonal analysis shows that surge risk is overwhelmingly a winter phenomenon. In winter, surge ratios exceed 1.5 at 30\% penetration even for 1–2 h outages and surpass 2.0 for longer events, whereas in spring, summer, and fall they remain an order of magnitude lower. These results emphasize the threshold-driven and temperature-contingent nature of HP inudced surges, with auxiliary resistance heating acting as a nonlinear amplifier, and underscore that resilience planning for electrified heating must explicitly account for winter restorations.

\begin{figure}[t]
\centering
\includegraphics[width=1.0\textwidth]{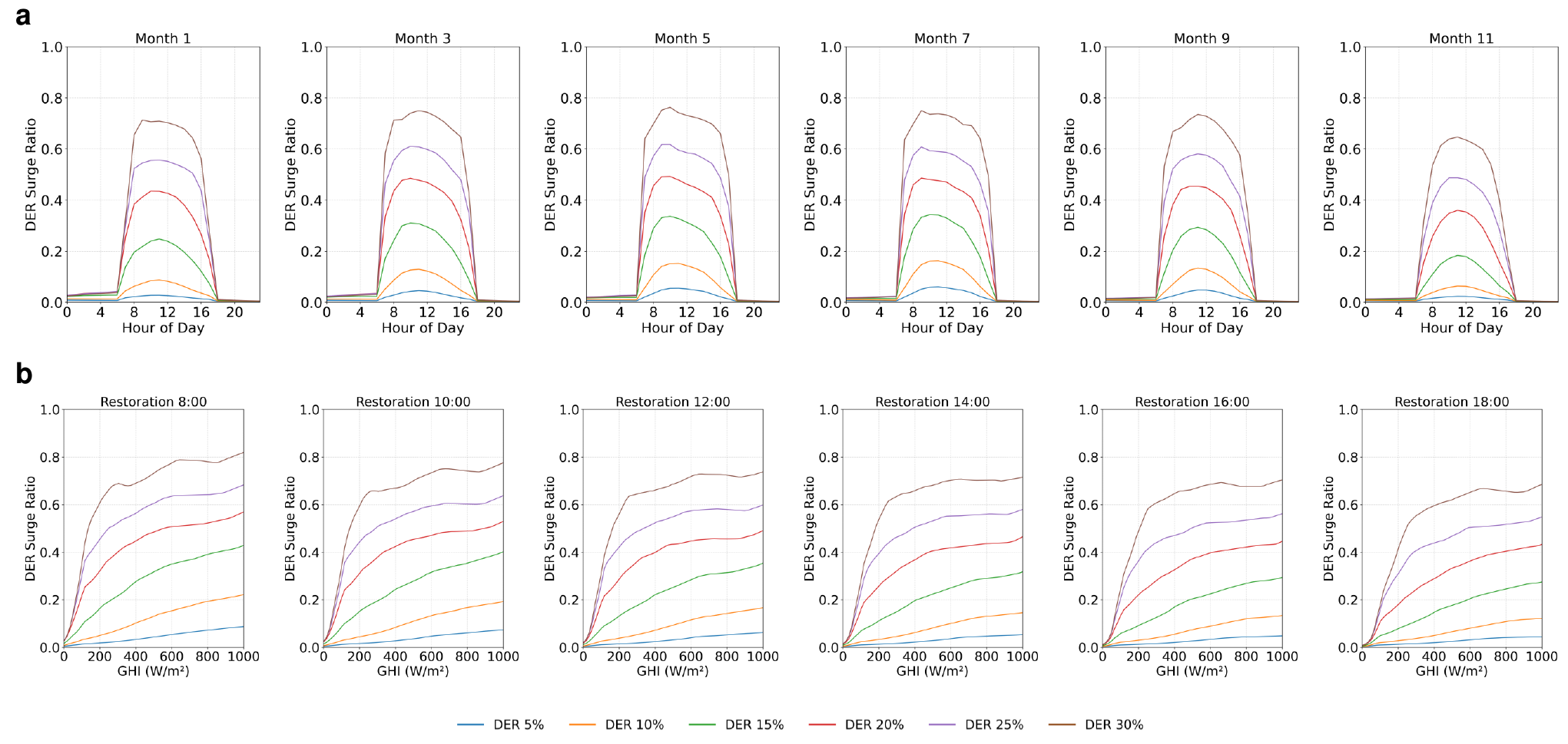}
\caption{Parametric analyses of DER-induced surge ratios under varying solar irradiance and restoration timing. (a) DER surge ratios as functions of hour of day across six representative months (Jan, Mar, May, Jul, Sep, and Nov). (b) DER surge ratios as functions of global horizontal irradiance (0–1000 W/m²) across six representative restoration hours (08:00, 10:00, 12:00, 14:00, 16:00, and 18:00). }\label{DEREST}
\end{figure}

In contrast to EV and HP surges, driven by deferred demand and thermodynamic recovery, respectively, DER induced surges originate from the delayed reconnection of PV inverters following grid restoration (Fig. \ref{DEREST}). The model indicates that surge ratios increase monotonically with penetration but exhibit distinctive temporal and irradiance-dependent patterns. Across the diurnal cycle, surges occur almost exclusively during daylight hours, rising rapidly after sunrise, peaking in the late morning to early afternoon, and declining toward sunset. The largest surges do not occur at solar noon but rather during the morning ramp, when solar irradiance is increasing most steeply. This indicates that DER-induced surges are governed by the rate of change of solar output rather than its absolute level. Restoration during these upward transitions amplifies the discrepancy between the baseline load and the absence of renewable generation. By comparison, midday restorations coincide with the irradiance plateau, where the surge is attenuated relative to the steep morning ramp, though still substantial. Irradiance magnitude further modulates outcomes: at fixed restoration times, surge ratios increase almost linearly with irradiance through moderate levels but show diminishing growth at higher values. For example, at 30\% penetration, surge ratios rise to 0.6–0.8 by 600 W/m², with smaller marginal increases thereafter. Unlike EVs and HPs, which accumulate unmet demand during outages, DER surges reflect delayed inverter reconnection that mirrors the solar profile. Consequently, seasonal differences are comparatively muted.

\section*{Projected load surge growth and mitigation}

The Indianapolis service territory, serving roughly 520,000 customers across 1,800 feeders, recorded a 2024 system peak demand of 2.80 GW, with headroom varying over the load profile and contracting to a minimum of about 0.50 GW at peak conditions. To project how post-outage surges may evolve under the  electrification and decarbonization transition, we re-simulate historical outage records as templates for a city-wide blackout restoration scenario under counterfactual 2035 adoption trajectories. In these projections, system-level covariates, including ambient temperature, restoration timing, outage duration, and solar irradiance, are fixed within each scenario to ensure physical consistency, while event-level heterogeneity in baseline demand and penetrations of EVs, HPs, and DERs is retained. We consider two prospective 2035 trajectories: (i) A baseline trajectory, derived by linearly extrapolating recent annual growth rates, yields penetration rates of approximately 10\% EVs, 30\% HPs, and 10\% DERs. (ii) A policy-aligned trajectory, consistent with U.S. national targets, reaching 30\% EVs, 40\% HPs, and 25\% DERs by 2035. To represent the scale of simultaneous restoration, we introduce the parameter $\alpha$, defined as the fraction of system baseline load re-energized within the first 15 minutes of recovery. Larger values of  $\alpha$ correspond to more aggressive restoration strategies that restore service quickly but risk higher surges, whereas smaller values imply more gradual recovery that mitigates surges at the expense of longer outage duration. For each $\alpha$, outage templates are repeatedly sampled until their combined baseline demand equals the specified fraction of the system load. Each draw yields an aggregate surge estimate, and the resulting ensemble produces a distribution from which we report the median, 95th percentile, and exceedance probability relative to restoration headroom (Method).

\begin{figure}[t]
	\centering
	\includegraphics[width=1.0\textwidth]{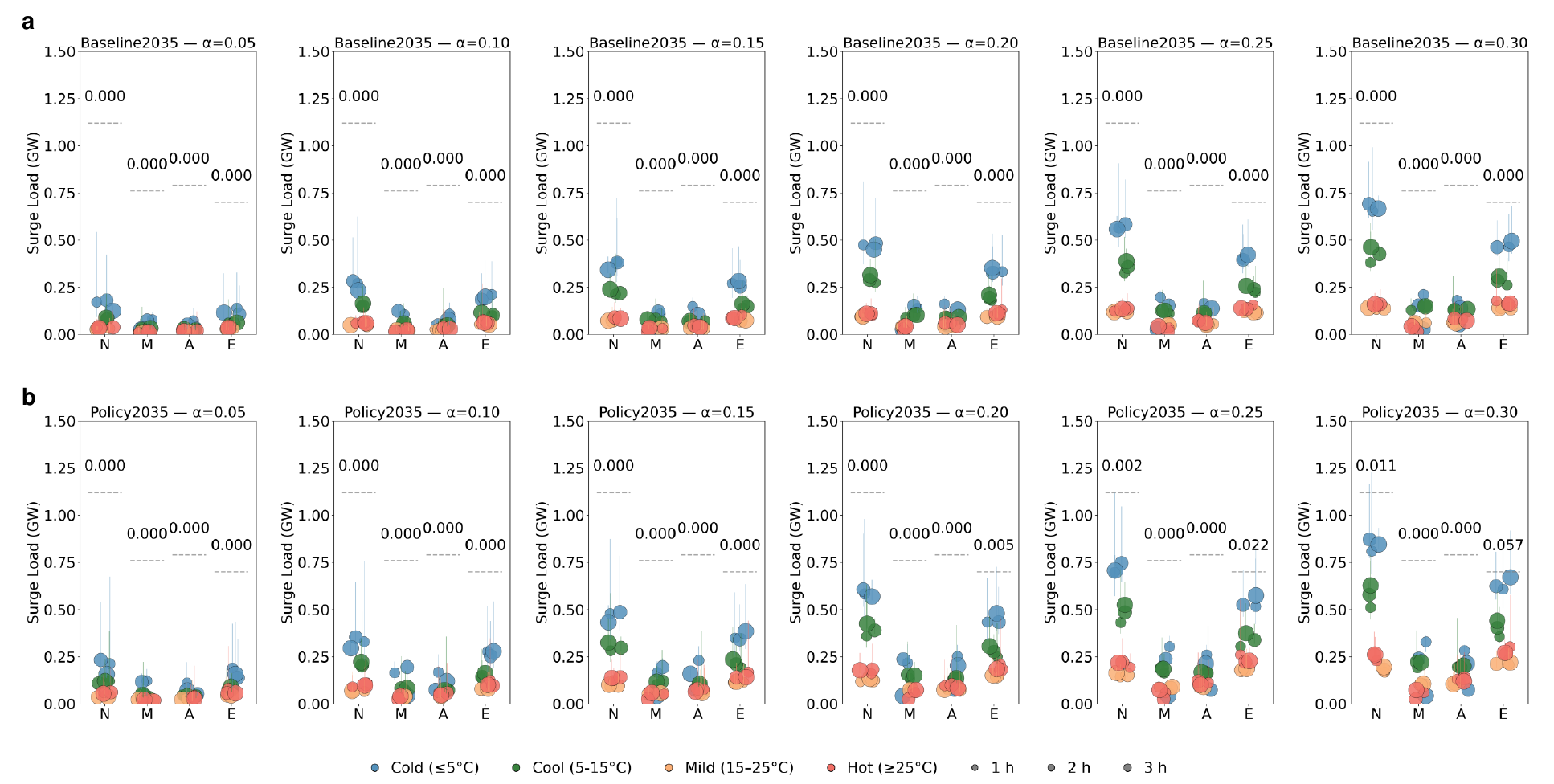}
	\caption{\textbf{City-wide surge as a function of restoration scale $\alpha$ and restoration window.} The gray dashed line denotes available headroom for each window; the number printed above this line is the exceedance probability. Points show mean surge load (GW); vertical bars indicate 95\% Monte Carlo intervals. Point color represents temperature bin and point size denotes outage duration. X-axis labels are N, M, A, E corresponding to Night, Morning, Afternoon, and Evening, respectively.}\label{Resmul}
\end{figure}

Across all simulations, post-outage surge magnitudes increase monotonically with the restoration scale $\alpha$, reflecting the growing challenge of re-energizing a larger share of city-wide demand in a short time (Fig. \ref{Resmul}). This simultaneity intensifies delayed DER reconnection and concentrates deferred EV charging and HP recovery. Similar monotonic patterns hold for ambient temperature and outage duration: for example, under the baseline trajectory, cold night events yield mean surges of 0.86–0.90 GW compared with just 0.40–0.43 GW under cool conditions, and 3 h outages amplify surges by 40\% relative to 1 h interruptions. Tail risk is governed by the restoration scale $\alpha$ and restoration window. Under the baseline trajectory, the probability of exceeding headroom is effectively zero across windows even at $\alpha=0.3$. Under the  policy-aligned trajectory, exceedance emerges as a low-probability tail concentrated in the evening, with a secondary contribution at night: at $\alpha=0.25$ the probabilities are 0.022 (evening) and 0.002 (night); and at $\alpha=0.30$ they increase to 0.057 (evening) and 0.011 (night). This concentration reflects limited evening headroom coincident with peak demand, stacked EV restarts, and cold-weather strip-heat activation. By contrast, although nighttime surges under extreme cold can be relatively larger, the greater off-peak headroom typically accommodates them. Morning and afternoon restorations remain comparatively manageable, with average surges rarely exceeding 0.3 GW. These observations highlight that evening restorations are the binding stress case, and a conservative operational limit of $\alpha\leq 0.15$ avoids exceedance under the policy-aligned 2035 pathway.

To counter these load surges, we evaluate staggered EV restarting, brief thermostat offsets, and accelerated DER reconnection (Methods). Together, these measures substantially suppress tail risk and narrow dispersion (Table \ref{FTB3}). Under the policy-aligned trajectory, all windows exhibit zero exceedance at $\alpha=0.20$; at $\alpha=0.25$, night remains at zero while the evening window declines to an exceedance probability of 0.004; and at $\alpha=0.30$, night remains at zero while the evening window drops from 0.057 to 0.019. Consistently, mean surges decrease and upper percentiles contract. For example, the evening mean declines from $0.61$–$0.67$ GW to $0.55$–$0.62$ GW, and the 3 h upper percentile drops from $0.92$ GW to $0.81$ GW. Nighttime means also fall from 0.81–0.85 GW to 0.73–0.80 GW, while morning and afternoon remain low at 0.19–0.26 GW. These results indicate that mitigation strategies effectively expands the safe restoration region: $\alpha=0.20$ constitutes a conservative operational guideline; $\alpha=0.25$ is broadly safe across restoration windows, and $\alpha=0.30$ is operationally acceptable, though a reduced evening tail risk remains.

\begin{table}[h]
	\caption{City‐wide load surge under 2035 trajectories with mitigation strategies}\label{FTB3}
	\begin{tabular*}{\textwidth}{@{\extracolsep\fill}l ccc cc ccc cc}
		\toprule
		& \multicolumn{5}{c}{\shortstack{Baseline trajectory\\(10\% EV, 30\% HP, 10\% DER)}}
		& \multicolumn{5}{c}{\shortstack{Policy-aligned trajectory\\(30\% EV, 40\% HP, 25\% DER)}}\\
		\cmidrule(lr){2-6}\cmidrule(lr){7-11}
		\multirow{2}{*}{Restoration Window}
		& \multicolumn{3}{c}{Surge (GW)} & \multicolumn{2}{c}{Exceed. Prob.}
		& \multicolumn{3}{c}{Surge (GW)} & \multicolumn{2}{c}{Exceed. Prob.}\\
		& 1 h & 2 h & 3 h & $\alpha=0.25$ & $\alpha=0.30$
		& 1 h & 2 h & 3 h & $\alpha=0.25$ & $\alpha=0.30$\\
		\midrule
		\multirow{3}{*}{Night}
		& 0.53 & 0.58 & 0.61 & \multirow{3}{*}{0.000} & \multirow{3}{*}{0.000}
		& 0.64 & 0.72 & 0.77 & \multirow{3}{*}{0.000} & \multirow{3}{*}{0.000}\\
		& (0.60) & (0.64) & (0.63) & & 
		& (0.73) & (0.80) & (0.78) & & \\
		& 0.85 & 0.81 & 0.68 & & 
		& 1.04 & 1.01 & 0.84 & & \\
		\midrule
		\multirow{3}{*}{Morning}
		& 0.16 & 0.13 & 0.14 & \multirow{3}{*}{0.000} & \multirow{3}{*}{0.000}
		& 0.25 & 0.19 & 0.20 & \multirow{3}{*}{0.000} & \multirow{3}{*}{0.000}\\
		& (0.17) & (0.15) & (0.15) & & 
		& (0.26) & (0.21) & (0.21) & & \\
		& 0.18 & 0.24 & 0.18 & & 
		& 0.27 & 0.33 & 0.27 & & \\
		\midrule
		\multirow{3}{*}{Afternoon}
		& 0.15 & 0.12 & 0.13 & \multirow{3}{*}{0.000} & \multirow{3}{*}{0.000}
		& 0.24 & 0.18 & 0.19 & \multirow{3}{*}{0.000} & \multirow{3}{*}{0.000}\\
		& (0.15) & (0.13) & (0.13) & & 
		& (0.24) & (0.19) & (0.20) & & \\
		& 0.15 & 0.15 & 0.15 & & 
		& 0.24 & 0.23 & 0.23 & & \\
		\midrule
		\multirow{3}{*}{Evening}
		& 0.37 & 0.40 & 0.41 & \multirow{3}{*}{0.000} & \multirow{3}{*}{0.000}
		& 0.48 & 0.53 & 0.55 & \multirow{3}{*}{0.004} & \multirow{3}{*}{0.019}\\
		& (0.43) & (0.42) & (0.46) & & 
		& (0.55) & (0.55) & (0.62) & & \\
		& 0.58 & 0.49 & 0.60 & & 
		& 0.73 & 0.64 & 0.81 & & \\
		\botrule
	\end{tabular*}
	\footnotetext{City-wide surge (GW) is reported at the restoration scale $\alpha$ yielding the largest mean. Each cell shows low (first row), mean (second row, in parentheses), and high (third row). Exceedance probabilities are shown separately for $\alpha=0.25$ and $\alpha=0.30$; other scales $\alpha \leq 0.20$ are omitted due to 0 value.}
\end{table}

\section*{Discussion and conclusions}

Electrification and decarbonization reshape post-outage recovery from a phenomenon dominated by conventional CLPU to one governed by the coupled behavior of EVs, HPs, and DERs. Using metropolitan-scale outage records linked with AMI and submeter data, we show that higher penetrations of these assets causally elevate post-restoration surge ratios and amplify tail risk in evening restorations. Under a 2035 policy-aligned pathway, exceedance remains negligible in the morning and afternoon but becomes non-zero in the evening (0.022 at $\alpha$=0.25 and 0.057 at $\alpha$=0.30) with a secondary contribution at night (0.002 and 0.011, respectively), reflecting limited evening headroom coincident with deferred EV charging and cold-weather strip-heat activation. These mechanisms explain why unconditional, system-wide comparisons understate risk, whereas analyses conditioned on timing, duration, and temperature reveal binding reliability constraints.

From an operational perspective, the restoration scale $\alpha$ is a controllable decision variable rather than a fixed boundary condition. Without mitigation, a conservative cap of $\alpha\leq0.15$ avoids exceedance under the 2035 policy-aligned pathway. With targeted mitigation, including probabilistic EV restarts, brief thermostat offsets, and accelerated DER reconnection, the surge distribution shifts inward and its upper tail narrows: evening mean surge falls from 0.61–0.67 GW to 0.55–0.62 GW, the 3 h upper percentile contracts from 0.92 GW to 0.81 GW, and evening exceedance declines from 0.057 to 0.019 at $\alpha$=0.30; nighttime means also drop from 0.81–0.85 GW to 0.73–0.80 GW. These findings directly translate into implications for practice: system operators and planners can (i) incorporate $\alpha$ as an explicit optimization variable in re-energization schedules, (ii) implement randomized EV restart windows within managed charging programs, (iii) authorize short, automatically reversed thermostat offsets during cold-weather restorations, and (iv) shorten minimum dead-times and randomize inverter reconnection, (v) design blackstart capability and reserve margins against the full distribution of surge outcomes, co-optimizing restoration with demand-side mitigation measures.

Several limitations qualify the external validity of our findings. First, results are calibrated to the Indianapolis end-use demand composition, climate, and headroom; while mechanisms generalize, quantitative thresholds will vary with local EV adoption and charging behavior, heating portfolios, feeder topologies, and DER interconnection standards. Second, the projections hold system-level covariates fixed within each scenario to ensure physical consistency. In practice, however, extreme weather events often compound multiple stressors, which could amplify variability and exacerbate tail risks. Third, demand-side interventions entail comfort and equity considerations; thermostat offsets should remain brief, opt-out, and accompanied by safeguards to protect consumers.

Future research should extend this framework beyond city-wide outage dynamics to the national scale of energy transitions. Restoration headroom and surge-risk constraints need to be integrated into capacity planning to ensure that national decarbonization targets remain operationally feasible at the local level. At the same time, socio-technical factors, such as participation elasticities, comfort tolerance, trust in automation, and perceptions of fairness, should be quantified through utility pilot programs and incorporated into models to capture both equity and reliability outcomes. Scenario analysis should also move toward ensembles of compounding hazards, e.g., extreme cold combined with prolonged repairs, or heatwaves coinciding with smoke-attenuated PV generation, translating these into context-specific restoration policies. Finally, cross-city comparisons across diverse conditions, including EV adoption and charging behaviors, DER interconnection and reconnection rules, heating technology mixes, and regional weather regimes, will be critical for mapping how quantitative thresholds shift.

\section*{Methods}

\subsection*{Metrics and normalization}

For event $e$, we aggregate AMI and submeter readings over the 15-min interval immediately after restoration to obtain total demand and EV/HP demand, $P_{e,\mathrm{tot}}^{\mathrm{res}}$, $P_{e,\mathrm{ev}}^{\mathrm{res}}$, and  $P_{e,\mathrm{hp}}^{\mathrm{res}}$. For DER, we model the reconnection delay $\tau$ as a truncated normal distribution $p(\tau)$ on $[\tau_{\min},\infty]$, and compute the demand surge resulted by DER missing by $
	P_{e,\mathrm{der}}^{\text{res}}-P_{e,\mathrm{der}}^{\text{base}} = C_e(\tau_0)-\int_{\tau_{\min}}^{\Delta T} C_{e}(\tau_0+\tau) \, p(\tau) \, d\tau$. To isolate the surge effect, the asset-induced incremental surge ratios $s_{e,i}$ are defined relative to the event’s total baseline load  $P_{e,\mathrm{tot}}^{\mathrm{base}}$ (equation (\ref{M1})). This normalization ensures that the contributions of individual assets, together with residual components, sum consistently to the total event surge ratio (equation (\ref{M2})). In addition, electrification and decarbonization metrics are normalized for changes in service population and system scale (equation (\ref{M3})). Event-level EV and HP penetration rates, $r_{e,\mathrm{ev}}$ and  $r_{e,\mathrm{hp}}$, are quantified as the ratio of the number of EV and heat pump  submeters, $C_e^{\text{ev}}$ and $C_e^{\text{hp}}$, within the affected area, to the total number of smart meters, $C_e^{\text{sm}}$. Event-level DER penetration rate  $r_{e,\mathrm{der}}$ is defined as the ratio of the installed DER power capacity within the affected area $P_e^{\text{der}}$, to its daily peak load, $P_e^{\text{pk}}$.
\begin{equation}\label{M1}
s_{e,i} =\frac{P_{e,i}^{\mathrm{res}} - P_{e,i}^{\mathrm{base}}}{P_{e,\mathrm{tot}}^{\mathrm{base}}}, i \in \{\mathrm{tot},\mathrm{ev}, \mathrm{hp}, \mathrm{der}\}
\end{equation}

\begin{equation}
s_{e,\mathrm{tot}} = s_{e,\mathrm{ev}}+s_{e,\mathrm{hp}}+s_{e,\mathrm{der}}+s_{e,\mathrm{oth}}
\end{equation}

\begin{equation}\label{M2}
r_{e,\mathrm{ev}} = {C_e^{\text{ev}}}/{C_e^{\text{sm}}},\; r_{e,\mathrm{hp}} = {C_e^{\text{hp}}}/{C_e^{\text{sm}}},\; r_{e,\mathrm{der}} = {P_e^{\text{der}}}/{P_e^{\text{pk}}}
\end{equation}

\subsection*{Causal inference}

We employ a causal forest, a machine learning estimator that learns treatment‐effect heterogeneity directly from data, to estimate how asset penetration causally impacts the surge ratio. The local treatment effect is
\[
\tau(z) = \frac{\partial}{\partial x} \, \mathbb{E}[Y_e \mid X_e = x, Z_e = z],
\]
where treatment $X_e$ is the asset penetration under event $e$, outcome $Y_e$ is the event-level surge ratio, and confounders $Z_e$ include ambient temperature, outage duration, number of customers affected, and time-of-day, with solar irradiance additionally included for DERs. The conditional effect $\tau(z)$ represents the marginal impact of a 100 percentage‐point increase in penetration when holding confounders fixed. The ATE is then obtained by averaging over the distribution of confounders,
\[
\text{ATE} = \mathbb{E}_{Z}[\tau(Z)] = \int \tau(z) \, dF_{Z}(z).
\]

To avoid overfitting, the causal forest employs cross-fitting, where separate subsamples are used for training and effect estimation. Each tree is trained on a bootstrap subsample of its training fold, learning partitions of the confounder space that maximize heterogeneity in treatment effects. Within each terminal leaf, local treatment effects are estimated by regressing outcomes on treatments, and leaf-wise predictions are aggregated across trees to yield conditional effects. Aggregating these local treatment effects produces the global ATE. Finally, we report global ATEs scaled to a 10 percentage‐point penetration increase ($0.1 \times$ATE), together with 95\% confidence intervals from the forest-level estimator.

\subsection*{Multi-task transformer}

For post-outage load estimation, we construct a training dataset based on the heterogeneous dataset. Specifically, the training set consists of feeder‐level outage events. Each event is characterized by restoration timestamp (time-of-day and outage duration), ambient temperature, and contemporaneous event-level penetration of EVs, HPs, and DERs. The supervised targets are the incremental surge ratios $s_{e,i}$ defined in equation~(\ref{M1}), with components corresponding to EV, HP, DER, and residual.

Our model comprises a shared Transformer encoder followed by four decoder heads. At each time step $t$, the $d$-dimensional input vector, including outage duration, ambient temperature, weather covariates (e.g., irradiance, precipitable water), event-level asset penetration rates, and calendar features (hour, day-of-week, month),
is linearly projected into a 
$D$-dimensional feature space and augmented with positional embeddings. This $T\times D$ sequence is passed through $L$ layers of multi‐head self-attention and position-wise feed‐forward sublayers, each employing residual connections, layer normalization, and dropout. The encoder’s output is global‐average-pooled across time to yield a single event embedding $\mathbb{R}^D$. Four task‐specific decoder multilayer perceptrons (MLPs) then map this embedding to predict incremental fractions $\widehat{s}_{e,i}$. Each MLP comprises a linear layer $\mathbb{R}^D\rightarrow \mathbb{R}^H$ with ReLU and dropout, followed by a linear $\mathbb{R}^H\rightarrow \mathbb{R}$ projection. 
All decoder heads and the shared encoder are trained jointly by minimizing the total loss defined as the sum of mean squared errors (MSE) between predicted and ground-truth incremental fractions.

\subsection*{Mitigation Strategy}

We propose three strategies to mitigate the growing post-outage load surge caused by rising electrification and decarbonization.

\medskip

\textbf{Staggered EV restarting.} Let \(M\) denote the number of EV chargers within the affected outage area, and let $P_{\mathrm{EV},m}(t)$ represent the power consumption of submeter $m$ at time $t$ immediately after restoration. To simulate staggered reconnection, we define a probabilistic restart window $[t_{1},t_{2}]$, over which we perform $K$ Monte Carlo trials. For each trial $k$, we independently sample a delay for each charger according to
$
    \tau_{m}^{(k)} \sim \mathcal{U}[t_{1},t_{2}]
$. For each charger 
$m$ and trial $k$, we construct a delayed post-restoration load profile (equation (\ref{EVMC1})). Aggregating across all EV chargers, we extract the EV load profile within the window $[t_{1},t_{2}]$, and compute the average load profile across  
$K$ trials.
\begin{equation}\label{EVMC1}
    P_{\mathrm{EV},m}^{(k)}(t)=
    \begin{cases}
      P_{\mathrm{EV},m}\bigl(t - \tau_{m}^{(k)}\bigr), & t \ge \tau_{m}^{(k)}\\
      0, & t < \tau_{m}^{(k)}
    \end{cases}
\end{equation}

In addition, the event-specific EV load surge is computed as the average value over the 15-minute interval (equation (\ref{EVMC2})).
Also, the resulting average reduction factor $\gamma_{EV}$, which is outage event-specific, is given by equation \ref{EVMC3}), where $P_{\mathrm{EV}}=
   \sum_{m=1}^{M} 
    P_{\mathrm{EV},m}$ represents the aggregated EV load of a specific event based on the submeter data.
\begin{equation}\label{EVMC2}
     P'_{\mathrm{EV}}
    =
    \frac{1}{15}\sum_{t=0}^{15\mathrm{min}}\Big\{\frac{1}{K} \sum_{k=1}^{K} 
    \sum_{m=1}^{M} 
    P_{\mathrm{EV},m}^{(k)}(t)\Big\}
\end{equation}
\begin{equation}\label{EVMC3}
    \gamma_{\mathrm{EV}}= 1-  P'_{\mathrm{EV}}/P_{\mathrm{EV}}
\end{equation}

\medskip
\textbf{Brief thermostat offsets.} To quantify the temperature sensitivity of heat pump power consumption, we apply a piecewise-linear regression using heat pump submeter. Specifically, we examine the relationship between the incremental heat pump surge ratio, $\Delta_{e,\mathrm{HP}}$, and the temperature difference $\Delta T=T_{\mathrm{out}}-T_{\mathrm{set}}$, where $T_{\mathrm{out}}$ denotes the outdoor ambient temperature and $T_{\mathrm{set}}$ denotes the indoor thermostat settings. Each data point in the resulting scatter plot corresponds to a single outage event, with the x-axis representing $\Delta T$ and the y-axis representing $\Delta_{e,\mathrm{HP}}$. The data are segmented into three temperature regimes. A piecewise-linear regression function $f_r$ is then fitted to capture three  operating regimes:
\begin{equation}
    f_r(\Delta T) = \begin{cases} \beta_{\text{cold}}  \Delta T + \alpha_{\text{cold}}, &  \Delta T < -5\,^\circ\text{C}  \\ \beta_{\text{mild}}  \Delta T + \alpha_{\text{mild}}, & |\Delta T| \leq 5\,^\circ\text{C} \\ \beta_{\text{hot}}  \Delta T + \alpha_{\text{hot}}, &  \Delta T > +5\,^\circ\text{C} \end{cases}
\end{equation}

The heat pump load surge reduction factor $\gamma_{HP}$ is given by
\begin{equation}\label{EVMC3}
    \gamma_{\mathrm{HP}}= 1-  f_r(\Delta T-\Delta T_{\mathrm{set}})/f_r(\Delta T)
\end{equation}

Note that in the cold regime, a negative thermostat offset ($\Delta T_{\mathrm{set}}<0$), corresponding to a heating setback, reduces heat pump power consumption. Similarly, in the hot regime, a positive offset ($\Delta T_{\mathrm{set}}>0$), corresponding to a cooling setback, also results in decreased power consumption.

\medskip

\textbf{Accelerated DER reconnection.}  We analyze a modified setting 
$\tau_{\min}'$ together with a randomized soft-start window, such that each inverter resynchronizes at a delay $t \sim p'(t)$ truncated to $[\tau_{\min}', 15 \text{min}]$. The missing DER power in the 15-min interval after reconnection acceleration is give by equation (\ref{ADR1}). Then, the DER load surge reduction factor $\gamma_{\mathrm{DER}}$ is defined as  the ratio of the missing DER power after acceleration to that before acceleration (equation (\ref{ADR2})).
\begin{equation}\label{ADR1}
    P_{\mathrm{DER}}^{'} = \frac{1}{15}\sum_{t=0}^{15\mathrm{min}}\Big\{C_e(\tau_0)-\int_{\tau_{\min}'}^{15\mathrm{min}} C_{e}(\tau_0+t) p'(t)  dt\Big\}
\end{equation}
\begin{equation}\label{ADR2}
    \gamma_{\mathrm{DER}} = P_{\mathrm{DER}}'/P_{\mathrm{DER}} 
\end{equation}

\bibliography{sn-bibliography}

\begin{thebibliography}{10}
\expandafter\ifx\csname url\endcsname\relax
  \def\url#1{\burl{#1}}\fi
\expandafter\ifx\csname urlprefix\endcsname\relax\def\urlprefix{URL }\fi
\providecommand{\bibinfo}[2]{#2}
\providecommand{\eprint}[2][]{\url{#2}}
\providecommand{\doi}[1]{\url{https://doi.org/#1}}
\bibcommenthead

\bibitem{IEAEV}
\bibinfo{author}{{IEA}}.
\newblock \bibinfo{title}{Global ev outlook 2024} (\bibinfo{year}{2024}).
\newblock \urlprefix\url{https://www.iea.org/reports/global-ev-outlook-2024}.

\bibitem{IEAHP}
\bibinfo{author}{{IEA}}.
\newblock \bibinfo{title}{The future of heat pumps} (\bibinfo{year}{2022}).
\newblock \urlprefix\url{https://www.iea.org/reports/the-future-of-heat-pumps}.

\bibitem{EUDER}
\bibinfo{author}{{European Commission}}.
\newblock \bibinfo{title}{Repowereu: Joint european action for more affordable,
  secure and sustainable energy} (\bibinfo{year}{2022}).
\newblock
  \urlprefix\url{https://commission.europa.eu/topics/energy/repowereu_en}.

\bibitem{woody2023decarbonization}
\bibinfo{author}{Woody, M.}, \bibinfo{author}{Keoleian, G.~A.} \&
  \bibinfo{author}{Vaishnav, P.}
\newblock \bibinfo{title}{Decarbonization potential of electrifying 50\% of us
  light-duty vehicle sales by 2030}.
\newblock \emph{\bibinfo{journal}{Nat. Commun.}} \textbf{\bibinfo{volume}{14}},
  \bibinfo{pages}{7077} (\bibinfo{year}{2023}).

\bibitem{rosenow2022heating}
\bibinfo{author}{Rosenow, J.}, \bibinfo{author}{Gibb, D.},
  \bibinfo{author}{Nowak, T.} \& \bibinfo{author}{Lowes, R.}
\newblock \bibinfo{title}{Heating up the global heat pump market}.
\newblock \emph{\bibinfo{journal}{Nat. Energy}} \textbf{\bibinfo{volume}{7}},
  \bibinfo{pages}{901--904} (\bibinfo{year}{2022}).

\bibitem{zhao2024impacts}
\bibinfo{author}{Zhao, J.}, \bibinfo{author}{Li, F.} \& \bibinfo{author}{Zhang,
  Q.}
\newblock \bibinfo{title}{Impacts of renewable energy resources on the weather
  vulnerability of power systems}.
\newblock \emph{\bibinfo{journal}{Nat. Energy}} \bibinfo{pages}{1--8}
  (\bibinfo{year}{2024}).

\bibitem{wang2023sequential}
\bibinfo{author}{Wang, Y.} \emph{et~al.}
\newblock \bibinfo{title}{Sequential load restoration with soft open points and
  time-dependent cold load pickup for resilient distribution systems}.
\newblock \emph{\bibinfo{journal}{IEEE Trans. on Smart Grid}}
  \textbf{\bibinfo{volume}{14}}, \bibinfo{pages}{3427--3438}
  (\bibinfo{year}{2023}).

\bibitem{li2024impact}
\bibinfo{author}{Li, Y.} \& \bibinfo{author}{Jenn, A.}
\newblock \bibinfo{title}{Impact of electric vehicle charging demand on power
  distribution grid congestion}.
\newblock \emph{\bibinfo{journal}{Proceedings of the National Academy of
  SciencesProc. Natl. Acad. Sci. U.S.A.}} \textbf{\bibinfo{volume}{121}},
  \bibinfo{pages}{e2317599121} (\bibinfo{year}{2024}).

\bibitem{NRELHP}
\bibinfo{author}{{National Renewable Energy Laboratory, USA}}.
\newblock \bibinfo{title}{Field validation of air-source heat pumps for cold
  climates} (\bibinfo{year}{2023}).
\newblock \urlprefix\url{https://docs.nrel.gov/docs/fy23osti/84745.pdf}.

\bibitem{photovoltaics2018ieee}
\bibinfo{author}{Photovoltaics, D.~G.} \& \bibinfo{author}{Storage, E.}
\newblock \bibinfo{title}{Ieee standard for interconnection and
  interoperability of distributed energy resources with associated electric
  power systems interfaces}.
\newblock \emph{\bibinfo{journal}{IEEE Std}} \textbf{\bibinfo{volume}{1547}},
  \bibinfo{pages}{2018} (\bibinfo{year}{2018}).

\bibitem{kulkarni2024enhancing}
\bibinfo{author}{Kulkarni, M.~S.} \emph{et~al.}
\newblock \bibinfo{title}{Enhancing grid resiliency in distributed energy
  systems through a comprehensive review and comparative analysis of islanding
  detection methods}.
\newblock \emph{\bibinfo{journal}{Sci. Rep.}} \textbf{\bibinfo{volume}{14}},
  \bibinfo{pages}{12124} (\bibinfo{year}{2024}).

\bibitem{xu2025quantifying}
\bibinfo{author}{Xu, L.}, \bibinfo{author}{Lin, N.}, \bibinfo{author}{Poor,
  H.~V.}, \bibinfo{author}{Xi, D.} \& \bibinfo{author}{Perera, A.}
\newblock \bibinfo{title}{Quantifying cascading power outages during climate
  extremes considering renewable energy integration}.
\newblock \emph{\bibinfo{journal}{Nat. Commun.}} \textbf{\bibinfo{volume}{16}},
  \bibinfo{pages}{2582} (\bibinfo{year}{2025}).

\bibitem{lee2022community}
\bibinfo{author}{Lee, C.-C.}, \bibinfo{author}{Maron, M.} \&
  \bibinfo{author}{Mostafavi, A.}
\newblock \bibinfo{title}{Community-scale big data reveals disparate impacts of
  the texas winter storm of 2021 and its managed power outage}.
\newblock \emph{\bibinfo{journal}{Humanit. Soc. Sci. Commun.}}
  \textbf{\bibinfo{volume}{9}}, \bibinfo{pages}{1--12} (\bibinfo{year}{2022}).

\bibitem{feng2025hurricane}
\bibinfo{author}{Feng, K.} \emph{et~al.}
\newblock \bibinfo{title}{Hurricane ida’s blackout-heatwave compound risk in
  a changing climate}.
\newblock \emph{\bibinfo{journal}{Nat. Commun.}} \textbf{\bibinfo{volume}{16}},
  \bibinfo{pages}{1--10} (\bibinfo{year}{2025}).

\bibitem{feng2022tropical}
\bibinfo{author}{Feng, K.}, \bibinfo{author}{Ouyang, M.} \&
  \bibinfo{author}{Lin, N.}
\newblock \bibinfo{title}{Tropical cyclone-blackout-heatwave compound hazard
  resilience in a changing climate}.
\newblock \emph{\bibinfo{journal}{Nat. Commun.}} \textbf{\bibinfo{volume}{13}},
  \bibinfo{pages}{4421} (\bibinfo{year}{2022}).

\bibitem{wu2022fragmentation}
\bibinfo{author}{Wu, H.} \emph{et~al.}
\newblock \bibinfo{title}{Fragmentation of outage clusters during the recovery
  of power distribution grids}.
\newblock \emph{\bibinfo{journal}{Nat. Commun.}} \textbf{\bibinfo{volume}{13}},
  \bibinfo{pages}{7372} (\bibinfo{year}{2022}).

\bibitem{ji2016large}
\bibinfo{author}{Ji, C.} \emph{et~al.}
\newblock \bibinfo{title}{Large-scale data analysis of power grid resilience
  across multiple us service regions}.
\newblock \emph{\bibinfo{journal}{Nat. Energy}} \textbf{\bibinfo{volume}{1}},
  \bibinfo{pages}{1--8} (\bibinfo{year}{2016}).

\bibitem{jain2017data}
\bibinfo{author}{Jain, R.~K.}, \bibinfo{author}{Qin, J.} \&
  \bibinfo{author}{Rajagopal, R.}
\newblock \bibinfo{title}{Data-driven planning of distributed energy resources
  amidst socio-technical complexities}.
\newblock \emph{\bibinfo{journal}{Nat. Energy}} \textbf{\bibinfo{volume}{2}},
  \bibinfo{pages}{1--11} (\bibinfo{year}{2017}).

\bibitem{sturmer2024increasing}
\bibinfo{author}{St{\"u}rmer, J.} \emph{et~al.}
\newblock \bibinfo{title}{Increasing the resilience of the texas power grid
  against extreme storms by hardening critical lines}.
\newblock \emph{\bibinfo{journal}{Nat. Energy}} \textbf{\bibinfo{volume}{9}},
  \bibinfo{pages}{526--535} (\bibinfo{year}{2024}).

\bibitem{bennett2021extending}
\bibinfo{author}{Bennett, J.~A.} \emph{et~al.}
\newblock \bibinfo{title}{Extending energy system modelling to include extreme
  weather risks and application to hurricane events in puerto rico}.
\newblock \emph{\bibinfo{journal}{Nat. Energy}} \textbf{\bibinfo{volume}{6}},
  \bibinfo{pages}{240--249} (\bibinfo{year}{2021}).

\bibitem{choobdari2024robust}
\bibinfo{author}{Choobdari, M.}, \bibinfo{author}{Samiei~Moghaddam, M.},
  \bibinfo{author}{Davarzani, R.}, \bibinfo{author}{Azarfar, A.} \&
  \bibinfo{author}{Hoseinpour, H.}
\newblock \bibinfo{title}{Robust distribution networks reconfiguration
  considering the improvement of network resilience considering renewable
  energy resources}.
\newblock \emph{\bibinfo{journal}{Sci. Rep.}} \textbf{\bibinfo{volume}{14}},
  \bibinfo{pages}{23041} (\bibinfo{year}{2024}).

\bibitem{jacob2024real}
\bibinfo{author}{Jacob, R.~A.}, \bibinfo{author}{Paul, S.},
  \bibinfo{author}{Chowdhury, S.}, \bibinfo{author}{Gel, Y.~R.} \&
  \bibinfo{author}{Zhang, J.}
\newblock \bibinfo{title}{Real-time outage management in active distribution
  networks using reinforcement learning over graphs}.
\newblock \emph{\bibinfo{journal}{Nat. Commun.}} \textbf{\bibinfo{volume}{15}},
  \bibinfo{pages}{4766} (\bibinfo{year}{2024}).

\bibitem{AMIDOE}
\bibinfo{author}{{U.S. Department of Energy}}.
\newblock \bibinfo{title}{Advanced metering infrastructure and customer
  systems} (\bibinfo{year}{2016}).
\newblock
  \urlprefix\url{https://www.energy.gov/sites/prod/files/2016/12/f34/AMI%20Summary%20Report_09-26-16.pdf}.

\bibitem{FLISRDOE}
\bibinfo{author}{{U.S. Department of Energy}}.
\newblock \bibinfo{title}{Fault location, isolation, and service restoration
  technologies reduce outage impact and duration} (\bibinfo{year}{2009}).
\newblock
  \urlprefix\url{https://www.energy.gov/sites/prod/files/2016/12/f34/AMI%20Summary%20Report_09-26-16.pdf}.

\bibitem{NARRD}
\bibinfo{author}{{National Centers for Environmental Information, USA}}.
\newblock \bibinfo{title}{North american regional reanalysis dataset}
  (\bibinfo{year}{2023}).
\newblock \urlprefix\url{https://psl.noaa.gov/data/gridded/data.narr.html}.

\end{thebibliography}

\end{document}